# Sustained Amplification of Coherent Spin Waves by Parametric Pumping with Surface Acoustic Waves

Carson Rivard, Albrecht Jander and Pallavi Dhagat

School of Electrical Engineering and Computer Science, Oregon State University, Corvallis, Oregon, USA

**Abstract**

Parametric amplification offers a route to overcoming intrinsic damping in spin-wave systems, a key challenge in the development of magnonic signal processing and computing technologies. Here we demonstrate the sustained amplification of coherent forward volume magnetostatic spin waves in a yttrium-iron-garnet thin film using a traveling surface acoustic wave as a nonstationary pump. A gain of up to 6 dB is achieved under continuous pumping below the threshold for parametric instability. The interaction generates an idler wave at a distinct frequency, consistent with three-wave mixing governed by energy and momentum conservation. This approach enables stable, frequency- and wavevector-selective spin-wave gain using practical pump power levels, establishing acoustic wave pumping as a viable mechanism for realizing active components in integrated magnonic circuits.

**Introduction**

Spin waves—coherent, collective excitations of magnetization in ordered magnetic materials—are attracting increasing interest as information carriers for wave-based signal processing and computation devices.[1–4] Their quasi-particle analogues, known as magnons, represent the quantized energy and momentum carried by these waves. Practical applications of spin-wave-based or "magnonic" devices have, however, been limited by intrinsic magnetic damping, which causes rapid decay of spin-wave amplitude during propagation. Even in the lowest loss materials, based on yttrium-iron-garnet (YIG), the lifetime of magnons is at most a few microseconds.[5]

One promising approach to compensating spin-wave damping is parametric pumping, a nonlinear process in which energy from an external oscillating field, typically at twice the spin-wave frequency, is transferred to existing spin-wave modes. Parametric amplification of spin waves has been demonstrated in YIG [6–9] as well as other, higher-loss materials.[10,11] However, these demonstrations employed spatially uniform magnetic pumping—effectively a stationary, near-field electromagnetic drive—which excites a broad range of modes and often leads to parametric instability once the pump exceeds a critical amplitude. To suppress this instability, previous studies have employed pulsed pumping schemes to limit the duration of energy transfer. In contrast, this work employed a nonstationary, time- and space-periodic pump in the form of a traveling surface acoustic wave (SAW). This configuration provides wavevector-selective gain while remaining below the instability threshold, enabling continuous and stable amplification of coherent spin waves. In doing so, it addresses a key impediment to the realization of active magnonic devices.

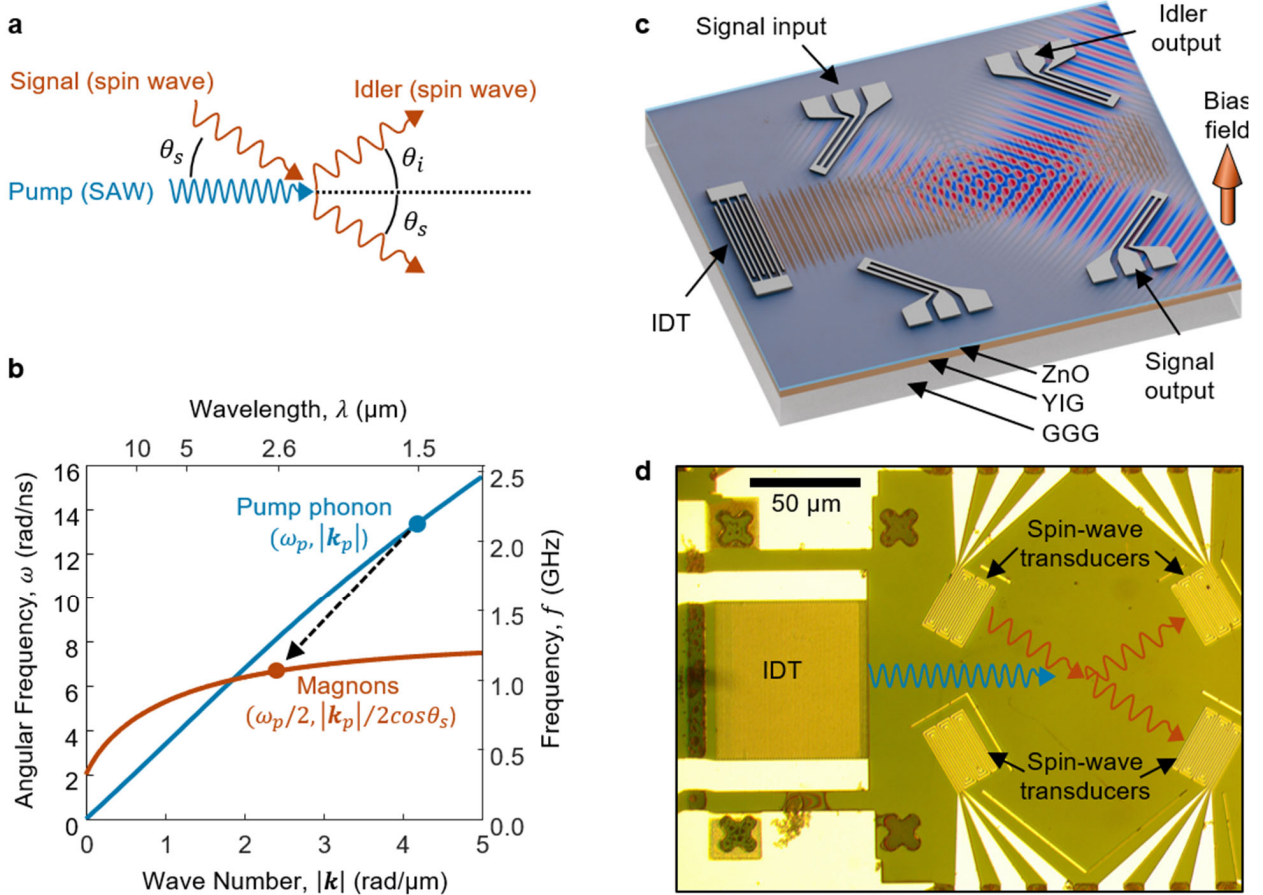

**Fig. 1 | Acoustic parametric pumping of spin waves. a.** Illustration of the parametric interaction between a spin wave and a SAW pump. **b.** Calculated dispersion curves for forward volume spin waves in a 1 μm-thick YIG film and surface acoustic waves in a layered medium of ZnO (425 nm) on YIG. In the parametric pumping interaction, a pump phonon splits into two magnons of half the pump frequency and the wave vectors must satisfy the conservation relation in equation (2). **c.** Experimental configuration. Spin waves are excited and detected in the YIG film by inductive transducers. The surface acoustic wave pump is generated using an interdigitated transducer (IDT) on piezoelectric ZnO. The bias magnetic field is oriented perpendicular to the film plane. **d.** Optical micrograph of a fabricate device.

## Acoustic parametric amplification of spin waves

The acoustic wave and spin wave interact via magnetoelasic coupling in YIG, by which the strain in the acoustic wave modulates the magnetic anisotropy. This time- and space-periodic modulation of the anisotropy can, under the conditions outlined below, result in the parametric amplification of the spin wave. In the process, energy and momentum are transferred from the acoustic wave to the spin wave. In the quasi-particle interpretation, the process is described as a phonon of the acoustic pump splitting into two lower-energy magnons, one with frequency and wave vector matching the existing "signal" spin wave and the other contributing to a secondary wave, conventionally called the "idler" wave. The idler wave must exist to satisfy the conservation of energy, $\hbar\omega$, and momentum, $\hbar \boldsymbol{k}$, in the three-wave mixing process. Specifically, the energy of the pump phonon lost must equal the sum of the energies of the two resulting magnons, giving the relation,

$$\omega_p = \omega_s + \omega_i ,  \quad\quad\quad (1)$$

in which $\omega_p$, $\omega_s$ and $\omega_i$ are the radian frequencies of the pump, signal and idler waves, respectively. Likewise, the momenta of the created magnons must add up to the momentum of the pump phonon lost, giving,

$$\boldsymbol{k}_p = \boldsymbol{k}_s + \boldsymbol{k}_i, \qquad (2)$$

where $\boldsymbol{k}_p$, $\boldsymbol{k}_s$ and $\boldsymbol{k}_i$ are the respective wave vectors.

Experimentally, we consider here only the degenerate case, where $\omega_s = \omega_i$ (i.e. the energies of the two magnons are the same) and the acoustic pump frequency is therefore twice the spin wave frequency. With a perpendicular magnetic bias field, the YIG film supports forward volume magnetostatic spin-wave modes with nearly isotropic in-plane propagation characteristics. Thus, for $\omega_s = \omega_i$ we also have $|\boldsymbol{k}_s| = |\boldsymbol{k}_i|$ and, to satisfy the vector sum in (2), the idler will travel away from the interaction region with angle[12]

$$\theta_i = -\theta_s, \qquad (3)$$

where $\theta_s$ is the angle of incidence of the signal wave with respect to the pump and $\theta_i$ is the angle at which the idler leaves the pumping region (see Fig. 1a). Accordingly, for degenerate pumping at a given angle of incidence, the magnitudes of the wave vectors must be related by

$$|\boldsymbol{k}_s| = |\boldsymbol{k}_i| = |\boldsymbol{k}_p|/2\cos(\theta_s). \qquad (4)$$

The parametric interaction can thus occur if the spin-wave dispersion curve is adjusted, by means of the bias field, to pass through the point $\omega_p/2$, $|\boldsymbol{k}_p|/2\cos(\theta_s)$, as illustrated in Fig. 1b.

As we have previously shown theoretically[13] and by micromagnetic modeling[12], the strength of the parametric interaction of forward volume waves pumped by SAWs depends on the angle of incidence, a consequence of the elliptical polarization of the spin waves and the phase difference between the longitudinal and perpendicular strain components of the SAW. The parametric coupling strength in the YIG film has a minimum around 20° and increases with increasing angle of incidence. Therefore, we use as large an angle as practical in the experiments, limited by the transducer linewidths that can be reliably patterned.

The device shown in Fig. 1d was fabricated to experimentally investigate the SAW pumping of spin waves. SAWs with a wavelength of 1.5 µm and a frequency of approximately 2 GHz are generated using an Al interdigitated transducer (IDT) on a thin film of piezoelectric ZnO. The acoustic waves penetrate the ZnO as well as the underlying 1-µm-thick YIG film (see Supplementary Information Section 4). Inductive spin wave transducers were arranged to launch a beam of spin waves to intersect the SAW an angle, $\theta_s = 30°$. The spin-wave transducers were designed for wavelengths of 2.6 µm in accordance with (4). The parametrically pumped signal spin wave is detected on the opposite side with an identical transducer. A third transducer was arranged to detect the idler spin wave at the symmetrical angle, $\theta_i = -30°$.

**Results**

To investigate the acoustic parametric amplifications of spin waves, the signal transmitted between the opposing transducers was measured using a vector network analyzer, as illustrated in Fig. 2a. The transmission spectra, $S_{21}$, without pumping and at two different pump power levels are plotted in Fig. 2b. The central peak, arising from the transmitted signal spin wave, grows with increasing pump power. (The transmission spectrum is displaced slightly to the right with increasing pump power as the acoustic wave shifts the spin-wave dispersion curve by reducing the average magnetization in the film.) The peak received signal power is plotted in Fig. 2c as a function of SAW pump power for different levels of signal input power. To better visualize the amplification, the same data are replotted in Fig. 2d to show the gain in signal power relative to the unpumped baseline. Clear gain is observed for pump powers exceeding 0 dBm, rising monotonically to a maximum gain of approximately 6 dB near 10 dBm pump power. Beyond this point, the gain begins to decline, likely due to higher-order nonlinear processes or the onset of parametric instabilities generating unwanted spin-wave modes that scatter energy away from the desired signal. The gain is independent of input signal power until the spin waves are driven into the nonlinear regime at -20 dBm.

The idler wave is detected at the third transducer using a spectrum analyzer as illustrated in Fig. 2e. For this measurement, the signal frequency was de-tuned from the degenerate condition by 1 kHz (0.0001%) so that the idler frequency is slightly different and can be distinguished from feedthrough of the input signal as seen in Fig. 2f. The signal observed at $f_s = \frac{f_p}{2} - 1$ kHz is due to electromagnetic feedthrough between the spin-wave transducers. The idler observed at $f_i = \frac{f_p}{2} + 1$ kHz is a new frequency generated in the parametric interaction consistent with (1). The detected idler power as a function of SAW power is plotted in Fig. 2g for varying levels of signal input power. In the absence of either signal input or pump, no idler is detected above the background noise level of -165 dBm. The idler peak appears only in the presence of both pump and signal inputs and increases proportionately with both input signal power and pump power (Fig. 2g), in accordance with the expected bilinear dependence for three-wave mixing[14]. Above 10 dBm pump power, the idler also shows deviation, again suggesting onset of higher-order effects or instability.

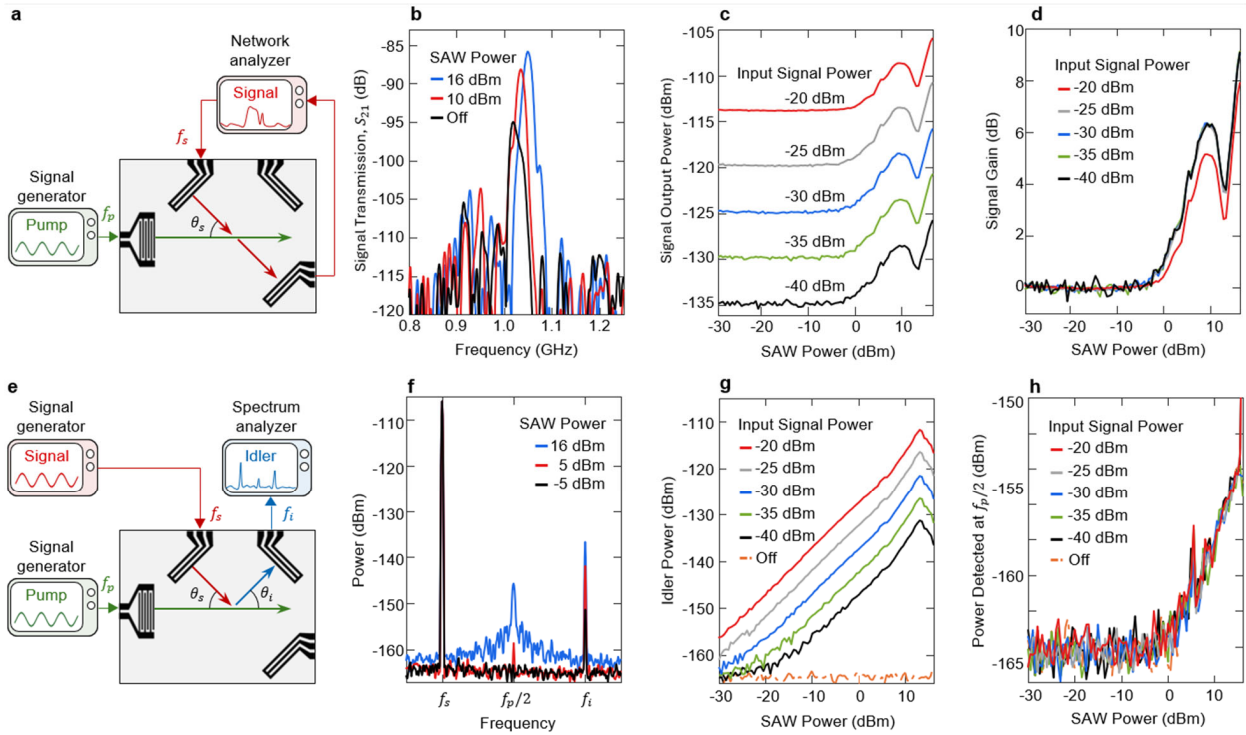

**Fig. 2 | Measurements of the parametric interaction between spin waves and surface acoustic waves. a.** Experimental setup for measuring signal transmission and gain. **b**. Transmission spectra from input to output transducer for various pump powers. **c.** Peak signal output power measured as a function of SAW power for different levels of input signal power. **d.** Gain in transmitted signal power normalized to the un-pumped transmission. **e.** Experimental setup for measuring idler power. **f.** Spectra observed at the idler output showing the generated idler signals at $f_i$ along with feedthrough of the signal frequency at $f_s$ and spin waves pumped from the thermal background at $f_p/2$. **g.** Detected idler power as a function of SAW pump power for different levels of input signal power. **h.** Signal detected at exactly $f_p/2$ resulting from pumping of spin waves from the thermal background.

In addition to the signal and idler, a third spectral component appears in Fig. 2f at exactly $\frac{f_p}{2}$, the degenerate condition. As seen in Fig. 2h, the power detected at this frequency is independent of the input signal power but grows with pump power. This signal is attributed to thermal magnon modes being parametrically amplified from the noise floor. This process occurs initially at the degenerate condition and broadens at higher pump power to include nearby frequencies (Fig. 2f).

Similar results, consistent with (1) and (4), were obtained with separate experimental devices having spin-wave transducers arranged at angles varying from 0° to 45° with respect to the SAW propagation (see Supplementary Information Section 1).

## Discussion

The experimental results summarized in Fig. 2 demonstrate sustained parametric amplification of coherent forward volume spin waves in a YIG film. The traveling SAW pump provides both temporal and spatial modulation, enabling frequency- and wavevector-selective coupling to spin-wave modes. This spatiotemporal selectivity allows continuous-wave operation without triggering parametric instability, a key limitation of conventional uniform magnetic field pumping. The detection of an idler wave at the expected frequency and angle, with bilinear power dependence, confirms the underlying three-wave mixing mechanism.

To evaluate whether the amplification truly overcomes intrinsic magnetic damping, we estimate the spin-wave decay length based on typical values of damping in these types of films. Using a Gilbert damping parameter of $\alpha = 0.0006$,[15] the decay length for 1.5 μm spin waves is about 780 μm.[16] Over the 140 μm propagation distance between the transducers, this corresponds to an intrinsic loss of about 1.6 dB. The observed amplification of up to 6 dB exceeds the loss, therefore providing a net gain of more than 4 dB. The SAW strain amplitude required to achieve this gain is estimated to be ~60 ppm (see Supplementary Information Section 4), which is well within the operating range of standard SAW devices.[17]

Beyond providing gain, parametric pumping with a traveling wave, when operating under nondegenerate conditions, enables spectral and spatial manipulation of spin-wave signals through the generation of an idler at frequencies and angles distinct from the input spin wave. This capability can support analog signal processing operations such as frequency translation[18], channel separation, directional control and nonreciprocal signal transmission.

In conclusion, this demonstration of directional, frequency-selective amplification under continuous pumping conditions establishes a new approach to spin-wave manipulation. The mechanism offers several advantages for the development of functional magnonic elements: it compensates intrinsic losses, preserves coherence, and permits frequency- and direction-resolved signal routing. As highlighted in prior studies,[2–4] such capabilities are essential for realizing spin-wave repeaters, logic gates, and active interconnects in wave-based signal processing and computing systems. The low power requirements and compatibility with established acoustic device structures further support integration into chip-scale circuits. Acoustic pumping thus offers a scalable and versatile tool for enabling coherent, active functionality in emerging magnonic technologies.

## Methods

### Transducer design

The SAW transducers are conventional IDTs using piezoelectric ZnO to mediate the electromechanical coupling.[19] The shortest achievable wavelength, and correspondingly highest frequency, of the SAW pump was limited by the linewidths that could be reliably patterned. The finger pitch of the IDTs was 0.75 µm (0.325 µm line and space) to obtain a SAW wavelength of 1.5 µm. With a calculated SAW phase velocity on the layered ZnO/YIG/GGG surface of around 3000 m/s this wavelength corresponds to a designed operating frequency near 2 GHz. The synchronous frequency of the fabricated transducer was measured to be 2048.2 GHz (see Supplementary Information Section 4). The IDTs have 50 finger pairs of 70 µm overlapping length to produce a 70 µm wide SAW beam.

The spin wave transducers are meandering coplanar ground-signal-ground waveguides with 6 meanders of 34 µm length to produce a 34 µm wide spin wave beam. The wavelength of the spin waves excited (or detected) equals twice the conductor-to-conductor spacing, designed to satisfy $\lambda_s = 2 \times 1.5$ µm $\cos(30°) = 1.3$ µm according to (4). The spacing loss due to the intervening ZnO results in relatively poor spin wave transduction efficiency with an estimated insertion loss per transducer of 30 dB. This loss was accepted over the additional fabrication complexity of removing the ZnO under the spin wave transducers.

### Device fabrication

The devices were fabricated on commercially available wafers (Innovent, E.V., Jena, Germany) consisting of a 1-µm-thick yttrium-iron-garnet (YIG) film grown by liquid-phase epitaxy on a <111>-oriented gadolinium gallium garnet (GGG) substrate. A 425 nm layer of zinc oxide (ZnO) was deposited on the YIG by magnetron sputtering. The ZnO thickness chosen is a trade-off between achieving good crystalline texture for piezoelectric coupling and ensuring sufficient penetration of SAW strain into the YIG film.

Interdigitated and meander-line transducers were patterned by reactive ion etching of a 160-nm-thick aluminum layer evaporated onto the ZnO, with a 5 nm titanium adhesion layer. Electron-beam lithography was used to define the fine transducer features, while optical lithography was employed for the larger probe pads. Scanning electron microscope images of the fabricated transducers are shown in Fig. 3.

### Bias field adjustment

The perpendicular magnetic bias field required to support forward volume spin waves was applied using a cylindrical (50 mm diameter, 25 mm thick) NdFeB permanent magnet surrounded with an electromagnet

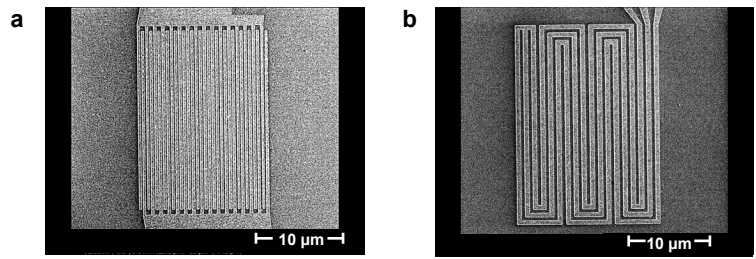

**Fig. 3 | Scanning electron microscope images of the fabricated transducers. a.** Interdigitated transducer for SAW generation. **b.** Meander line transducer for spin waves. (Probe pads not shown)

coil for tuning. For a given spin-wave excitation frequency, the magnetic field is adjusted to obtain a maximum in the transmission between the input and output transducers. At this field, the wavelength of the spin waves matches the periodicity of the meander-line transducer.

Due to the thin-film shape anisotropy, even a small angular deviation from the perpendicular orientation of the external bias field can lead to a much larger deviation of the internal field and magnetization angle. Consequently, the results are very sensitive to the precise alignment of the bias field which is adjusted by tilting the magnet assembly in the experimental setup. The perpendicular orientation is found by tuning the tilt until a minimum is obtained in the frequency spectrum of transmitted spin waves.[20]

*Measurements*

A vector network analyzer (Agilent E5071C) was used to measure the transmission and amplification of signal spin waves from the input transducer to the output transducer. The "time gating" feature[21] of the network analyzer was used to eliminate electromagnetic feedthrough of the signal between the cabling and transducers. Since the spin wave signal can be distinguished from the electromagnetic coupling by its longer transit time, a time gate of 50 ns can be used to eliminate the feedthrough.

The SAW is excited by driving the IDT at its synchronous frequency, $f_p$ = 2048.2 MHz, by a continuous sinusoidal signal from a microwave signal generator (Stanford Research Systems SG384).

For the idler measurements, the input spin wave transducer is driven with a continuous sinusoidal signal from a microwave signal generator (Agilent N5183A) while the idler output signal is observed using a spectrum analyzer (Tektronix RSA3308B). In this case, the idler signal is distinguished from the feedthrough by shifting the signal frequency down by 1 kHz relative to the degenerate case, i.e. $f_s$ = 1024.099 MHz instead of 1024.1 MHz. The resulting idler is, according to (1), shifted up by 1 kHz to 1024.101 MHz, where it can be separated from the feedthrough using a 7.8 Hz resolution bandwidth on the spectrum analyzer.

All electrical connections to the device were through 50 Ω coaxial cables and coplanar ground-signal-ground microwave probes. For both the network analyzer as well as the spectrum analyzer measurements, a low-noise amplifier (Mini Circuits ZKL-33ULN-S+, not shown in Fig. 2) is used to improve the signal-to-noise ratio.

*Dispersion calculations*

The dispersion curve for forward volume waves in the YIG film, shown in Fig. 1b, was calculated using the approximate formula derived by Kalinikos and Slavin, [22]

$$\omega^2 = \omega_0 \left[\omega_0 + \omega_M \left(1 - \frac{1-e^{-kd}}{kd}\right)\right],$$

for a film with thickness, $d$ = 1 μm, using $\omega_0 = -\gamma\mu_0 H_0$ and $\omega_M = -\gamma\mu_0 M_S$. We used standard values for the gyromagnetic ratio, $\gamma$ = -2π×28 GHz/T, vacuum magnetic permeability, $\mu_0$ = 1.257×10⁻⁶ H/m, and saturation magnetization, $M_S$ = 140 ×10³ A/m for YIG.[16]

The dispersion curve for the SAW traveling on the layered ZnO/YIG/GGG surface was calculated numerically using Green's function analysis[23] with the following material properties:[24] density, $\rho$ = 5.6 g/cm³, elastic constants, $C_{11}$ =209.7 GPa, $C_{12}$ =121.1 GPa, $C_{13}$ =105.1 GPa, $C_{33}$ =210.9 GPa, $C_{44}$ = 42.5 GPa for ZnO; $\rho$ = 5.17 g/cm³, $C_{11}$ = 269 GPa, $C_{22}$ = 108 GPa, $C_{44}$ = 76.4 GPa for YIG; and $\rho$ = 7.094 g/cm³, $C_{11}$ = 285.7 GPa, $C_{22}$ = 114.9 GPa, $C_{44}$ = 90.2 GPa for GGG.

# Supplementary Information

## to

## Sustained Amplification of Coherent Spin Waves by Parametric Pumping with Surface Acoustic Waves

### 1. Additional results for other wavelengths and angles of incidence

Several different experimental devices, designed to operate at different wavelengths and angles of incidence, were fabricated and evaluated. Table S1 lists the designed spin-wave and SAW wavelengths of four devices. The angle of incidence and the experimentally determined bias field requirement is also noted. In each case, the transducer wavelengths were designed to be consistent with equation (4).

A summary of the experimental results is shown in Fig. S1. Signal gain and idler generation were observed in all of the devices, confirming parametric pumping of the coherent input signal spin wave. Pumping of spin waves from the thermal background is observed in all the devices except for device #2. However, since only specific wave vectors are detected by the receiving transducers. Thermal spin waves with other wave vectors may have been pumped but are not detected.

*Table S1. Parameters of Devices Tested*

| Device # | SAW Pump Wavelength (µm) | SAW Pump Frequency (MHz) | Spin Wave Wavelength (µm) | Spin Wave Frequency (MHz) | Angle of Incidence | Bias Field (mT) |
|---|---|---|---|---|---|---|
| 1 | 1.5 | 2052.2 | 2.6 | 1026.1 | 30° | 179.3 |
| 2 | 2.0 | 1605.2 | 2.8 | 802.6 | 45° | 175.5 |
| 3 | 1.5 | 2045.2 | 3.0 | 1022.6 | 0° | 179.7 |
| 4 | 2.0 | 1602.2 | 4.0 | 801.1 | 0° | 176.5 |

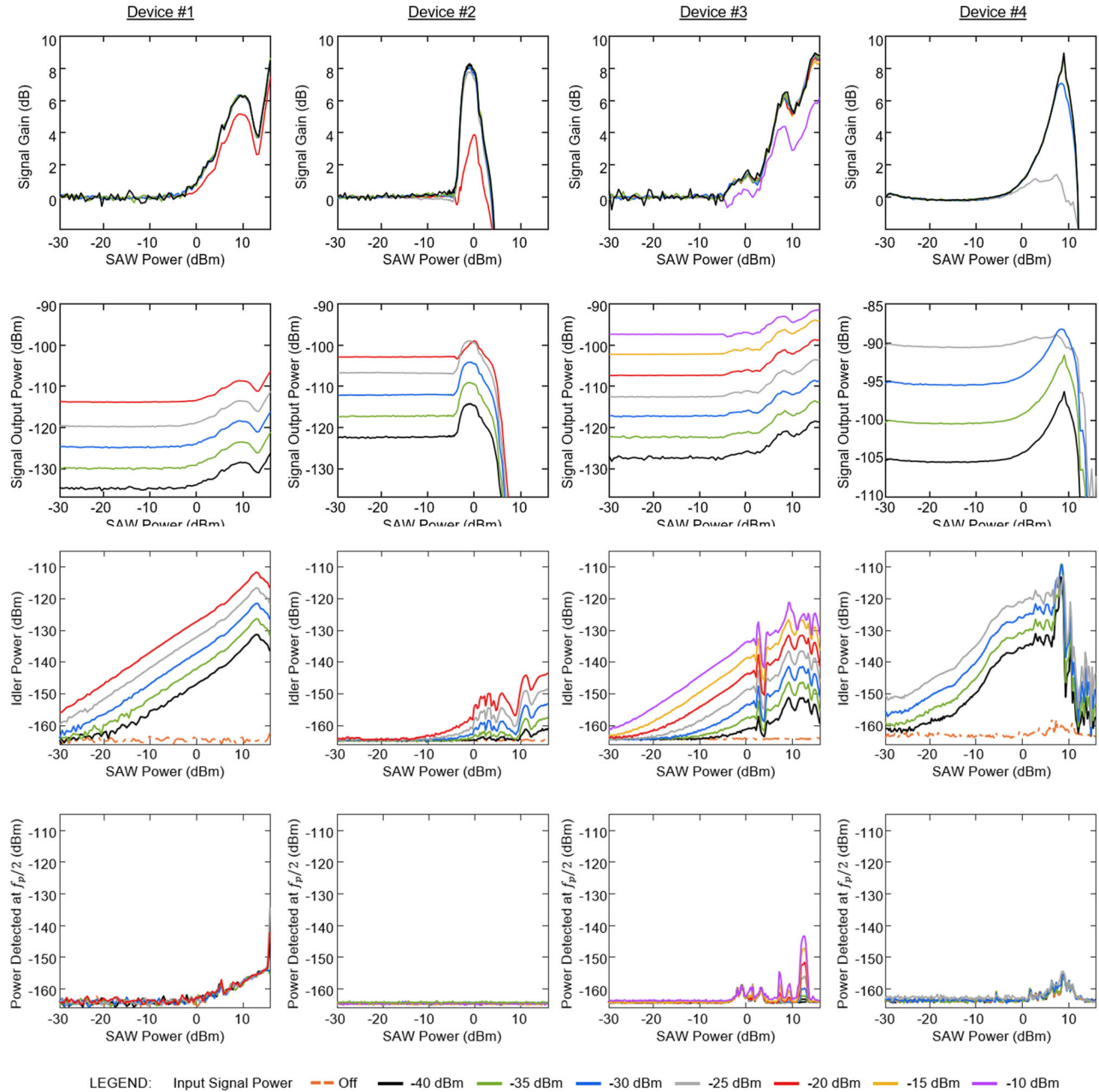

**Fig. S1 | Measurements as a function of SAW pump power for devices with different angles of incidence and pump wavelengths.** Columns correspond to devices listed in Table S1. **First row:** Signal gain relative to unpumped transmission. **Second row:** Peak signal output power. **Third row:** Detected idler power. **Fourth row:** Power detected at degenerate condition due to pumping from thermal background. Colored traces correspond to different input signal power as listed in the legend at the bottom.

## 2. "Null" experiment with reversed SAW propagation

The generation of the idler wave at a frequency distinct from any applied signal frequencies is a hallmark of the nonlinear parametric interaction. To eliminate the possibility that spurious nonlinearities in the instrumentation might be the source of the observed frequency mixing, we performed otherwise identical experiments with the direction of SAW propagation reversed in device #1. With the pump direction reversed, the vector relation in (2) is no longer satisfied and no idler should be generated.

The idler (red curve in Fig. S2a) observed in this case is diminished by about 20 dB, though not completely absent. The residual idler is likely due to a counter-propagating SAW partially reflected from the opposing IDT. As seen in Fig. S2b, negligible amplification of the signal spin wave is observed in the reverse pumping direction. Rather, a strong attenuation of the signal is seen above 10 dBm pump power, an effect attributed to the excitation of spin waves pumped from the thermal background leading to scattering of the signal spin wave. Pumping from the thermal background is independent of SAW direction.

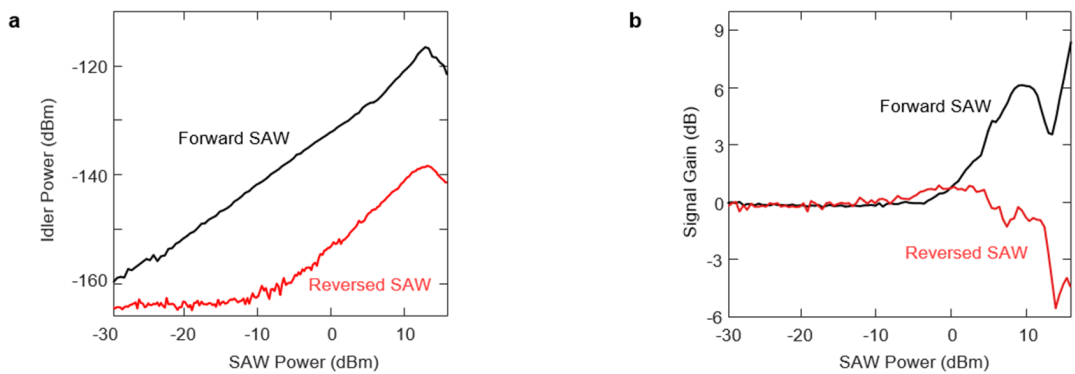

**Fig. S2 | Effect of reversing the SAW propagation direction.** Comparison of (**a**) idler generation and (**b**) signal amplification for forward (black) and reverse (red) direction of SAW propagation.

## 3. Identification of excited spin wave modes

Several different spin wave modes can propagate in the YIG film. To verify the conditions for excitation of the fundamental (n = 0) thickness mode of the forward volume waves, the transmission spectrum between the opposing spin wave transducers as a function of bias field is plotted in Fig. S3. This is compared to the calculated mode frequency corresponding to the 2.6 µm wavelength matching the transducer in Fig. 3. Plotting the power transmission at $f_p/2 = 1.025$ GHz, we verify that the n = 0 mode can propagate at a bias field near 179 mT.

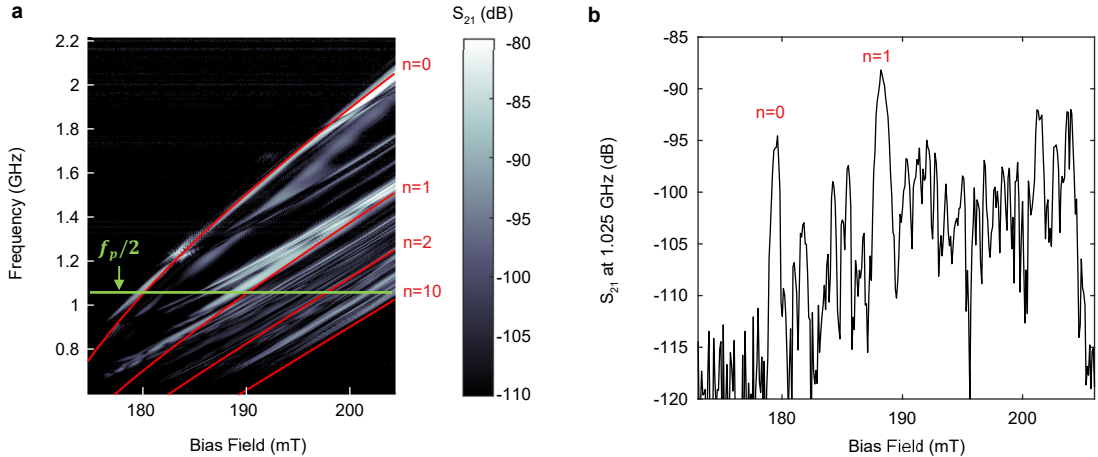

**Fig. S3 | Observed spin wave modes. a.** Transmission spectra, $S_{21}$, as a function of bias field measured between the opposing spin wave transducers in device #1. The horizontal green line shows the target frequency corresponding to ½ the pump frequency. The red lines are calculated frequencies for different thickness modes (n = 0, 1, 2…). **b.** Power transmission at $f_p/2 = 1.025$ GHz as a function of bias field (line cut along green line in **a.**).

## 4. Surface acoustic wave mode profile and strain amplitudes

The efficiency of acoustic wave generation by the IDT was estimated by comparing the electrical reflection coefficient, $S_{11}$, at and off the synchronous frequency, where the conversion from electrical to acoustic power is maximized. At resonance, the reflection loss is –1.21 dB (24% power absorption), compared to –0.51 dB (11% absorption) off resonance. Assuming the difference (13%) arises entirely from SAW transduction and that the acoustic power splits equally in both directions, an estimated 6.5% of the input power is carried by surface acoustic waves propagating to the right of the IDT.

At an input power of 0 dBm (1 mW), this corresponds to an acoustic power flow density of approximately 0.9 W/m within the 70 µm-wide SAW beam. The corresponding strain amplitude profile, resolved as a function of depth into the multilayer structure, is shown in Fig. S4. These results were obtained using open source MATLAB code,[25,26] which models SAWs in layered elastic media by numerically solving the Christoffel equation in each layer to extract eigenmodes and wavevectors. The Green's functions are then constructed to compute dispersion relations and depth-dependent strain fields.

The calculation shows that the SAW penetrates well into the YIG film with the peak of the longitudinal stain component, $\varepsilon_{xx}$, residing within the YIG and reaching a magnitude of about 20 ppm with an input power of 0 dBm. Since the strain amplitude is proportional to the square root of power, the 10 dBm threshold for instability in device #1 corresponds to roughly 60 ppm longitudinal strain amplitude.

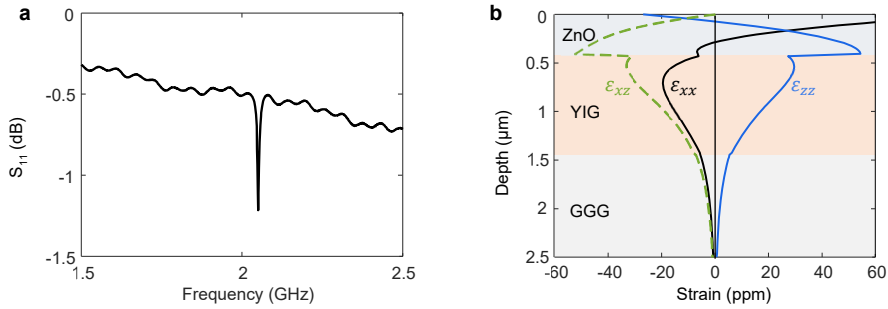

**Fig. S4 | Strain amplitude as a function of depth for the SAW**. **a.** Electrical power reflected from the IDT as a function of frequency around the synchronous frequency of 2.048 GHz. **b.** Longitudinal strain, $\varepsilon_{xx}$, perpendicular strain, $\varepsilon_{zz}$, and shear strain, $\varepsilon_{xz}$, components of the SAW as a function of depth into the layered structure of the device. Amplitudes are calculated for a drive of 0 dBm applied to the IDT.